\documentclass[9pt,twocolumn,twoside]{pnas-new}
\usepackage{bbold}

\templatetype{pnasresearcharticle} 

\begin{document}

\title{Microscopic Phase-Field Modeling}

\author[a,1]{Jaehyeok Jin}
\author[a,1]{David R. Reichman}

\affil[a]{Department of Chemistry, Columbia University, New York, New York 10027, United States}

\leadauthor{Jin}

\significancestatement{Rigorous ``bottom-up'' coarse-grained modeling has facilitated efficient simulations of complex systems, yet these methods have largely remained confined to the molecular regime. Our paper introduces a novel mesoscopic coarse-graining framework that enables the first-principles design of mesoscopic phase-field models, free of \textit{ad hoc} assumptions. Combined with coarse-grained enhanced sampling, our bottom-up phase-field modeling integrates atomistic details into a large-scale field-theoretic description, achieving a computational scale-up of over eight orders of magnitude. Because such approaches are lacking in conventional field-theoretic models, this framework represents a significant advancement in current systematic bottom-up methodologies for mesoscopic phase dynamics in complex materials, extending predictability beyond the reach of atomistic simulations and experimental probes.}

\authorcontributions{Author contributions: J.J. and D.R.R. designed research; J.J. and D.R.R. performed research; J.J. analyzed data; J.J. and D.R.R. wrote the paper.}
\authordeclaration{The authors declare no competing interest.}
\correspondingauthor{\textsuperscript{1}To whom correspondence should be addressed. E-mail: jj3296@columbia.edu or drr2103@columbia.edu}


\begin{abstract}Phase-field methods offer a versatile computational framework for simulating large-scale morphological evolution. However, the applicability and predictability of phase-field models are inherently limited by their \textit{ad hoc} nature, and there is currently no version of this approach that enables truly first-principles predictive modeling of large-scale non-equilibrium processes. Here, we present a bottom-up framework that provides a route to the construction of mesoscopic phase-field models entirely based on atomistic information. Leveraging molecular coarse-graining, we describe the formulation of an order parameter-based free energy functional appropriate for a phase-field description via the enhanced sampling of rare events.
We demonstrate our approach on ice nucleation dynamics, achieving a spatiotemporal scale-up of nearly $10^8$ times compared to the microscopic model. Our framework offers a unique approach for incorporating atomistic details into mesoscopic models and systematically bridges the gap between microscopic particle-based simulations and field-theoretic models. \end{abstract}

\dates{This manuscript was compiled on \today}
\doi{DOI to be provided by publisher}

\maketitle
\thispagestyle{firststyle}
\ifthenelse{\boolean{shortarticle}}{\ifthenelse{\boolean{singlecolumn}}{\abscontentformatted}{\abscontent}}{}

\firstpage[4]{5}

\dropcap{S}olidification and nucleation processes involving structural and dynamical transformations that couple atomistic-level microstructure with macroscopic properties are of pivotal importance in materials science and nanotechnology \cite{martin1997stability,altshuler2012mesoscopic}. However, explicating the mechanisms underlying these mesoscale processes, such as dendrite formation \cite{langer1980instabilities,david1992current} and heterogeneous nucleation \cite{kelton2010nucleation}, remains a frontier challenge. Molecular simulation offers an efficient means to elucidate microscopic factors that govern the small length scale processes underlying non-equilibrium growth processes \cite{Horstemeyer2010,shibuta2015solidification}. However, brute-force particle-based simulations face clear spatiotemporal limitations beyond the molecular level \cite{weinan2011principles}. 

To efficiently describe mesoscopic phase transformations, phase-field models are often used \cite{langer1986models, moelans2008introduction, provatas2011phase, steinbach2013phase}. Their free energy functionals are built in the spirit of the Ginzburg-Landau approach \cite{ginzburg2009theory}, a well-established framework for modeling phase transitions in materials and soft-matter physics \cite{aranson2002world}. In most practical phase-field formulations, however, the specific form of the free energy density is chosen phenomenologically and hence lacks a clear connection to the underlying microscopic physics \cite{plapp2011remarks}. Given that complex structures at large length scales are ultimately connected to the underlying microscopic interactions between atoms and molecules, such approaches may encounter challenges in accurately characterizing and predicting material growth across various spatiotemporal scales. This issue is crucial not only for understanding solidification through the symmetry and interactions of constituent molecules \cite{nelson2002defects}, but also for designing novel materials, where microscopic interactions determine macroscopic properties \cite{wittkowski2011microscopic}. Deriving a microscopic phase-field model requires scaling from the atomistic level to the larger mesoscopic level directly, but this process involves condensing millions of degrees of freedom into a compact set of parameters, a problem for which no general systematic framework currently exists.

To address this challenge, we develop a fully microscopic framework that enables the full parametrization of phase-field models from atomistic information by leveraging bottom-up coarse-graining (CG) and enhanced sampling techniques. Starting from the atomistic scale, we systematically transition to an intermediate molecular system through molecular coarse-graining. Enhanced sampling is then applied at this coarser level in a hierarchical manner to derive the corresponding phase-field representation. 

\section*{Conventional Phase-Field Model}

In modeling solidification, the free energy functional of a phase-field model $\mathcal{F}$ is typically expressed in terms of a single phase-field $\phi$ that takes values from 0 (liquid) to 1 (solid) to describe two-phase systems of volume $V$ \cite{boettinger2002phase,takaki2014phase,granasy2019phase}

\begin{equation} \label{eq:1}
    \mathcal{F}=\int_{V} dV (f_{chem}+f_{doub}+f_{grad}),
\end{equation}
where the chemical free energy density 
($f_{chem}$) interpolates between solid ($f_s$) and liquid ($f_l$) free energies using an \textit{ad hoc} interpolating polynomial $\bar{p}(\phi)$: $f_{chem}=\bar{p}(\phi)f_s +(1-\bar{p}(\phi))f_l$. Conventionally, $\bar{p}(\phi)=\phi^3(10-15\phi+6\phi^2)$ is adopted regardless of the system \cite{langer1986models}, as this form phenomenologically distinguishes $\phi=0$ and $1$ satisfying $\bar{p}(0)=0$ and $\bar{p}(1)=1$, in which the two phases are locally stable, i.e., $\bar{p}'(0)=\bar{p}'(1)=0$. The double-well potential imposes the free energy barrier $f_w$ at the interface as $f_{doub}=f_w\bar{q}(\phi)$ with a double-well interpolating polynomial $\bar{q}(\phi)=\phi^2(1-\phi)^2$. Lastly, the gradient free energy density $f_{grad}$ penalizes a sharp interface with a gradient coefficient $f_\epsilon$: $f_{grad}=f_\epsilon^2/2|\nabla \phi|^2$. 

The phenomenological description of Eq. (\ref{eq:1}) can distinguish between liquid ($\phi=0$) from solid ($\phi=1$) and is suitable for modeling various types of phase-evolution dynamics \cite{boettinger2002phase,takaki2014phase,granasy2019phase}. However, the accuracy and predictive capability of such \textit{ad hoc} approximations are questionable, as the free energy densities in Eq. (\ref{eq:1}) are typically drawn from experimental databases rather than derived from first principles \cite{saunders1998calphad}. While several studies have attempted to compute free energies from smaller-scale molecular simulations \cite{hoyt2001method,wu2006ginzburg,fu2017bridging,kavousi2019combined}, these efforts remain largely fragmented, relying on independent simulations to estimate different components of the free energy. Consequently, there is currently no systematic framework capable of determining all the parameters in Eq. (\ref{eq:1}) in a unified and self-contained manner. In particular, the functional forms of the exact polynomials ${p}(\phi)$ and ${q}(\phi)$ are intrinsically coupled to the underlying free energy terms, making it impractical to determine both the free energy densities and interpolating polynomials simultaneously from microscopic information. Instead, existing literature often employs the same \textit{ad hoc} polynomials $\bar{p}$ and $\bar{q}$ across chemically distinct systems without considering the system-specific dependencies \cite{plapp2011remarks}. This limitation significantly restricts the predictive power of current phase-field models and serves as the main motivation for the present work.

\section*{Microscopic Framework}

\subsection*{Hierarchical Coarse-Graining}
\begin{figure}
\includegraphics[width=1\columnwidth]{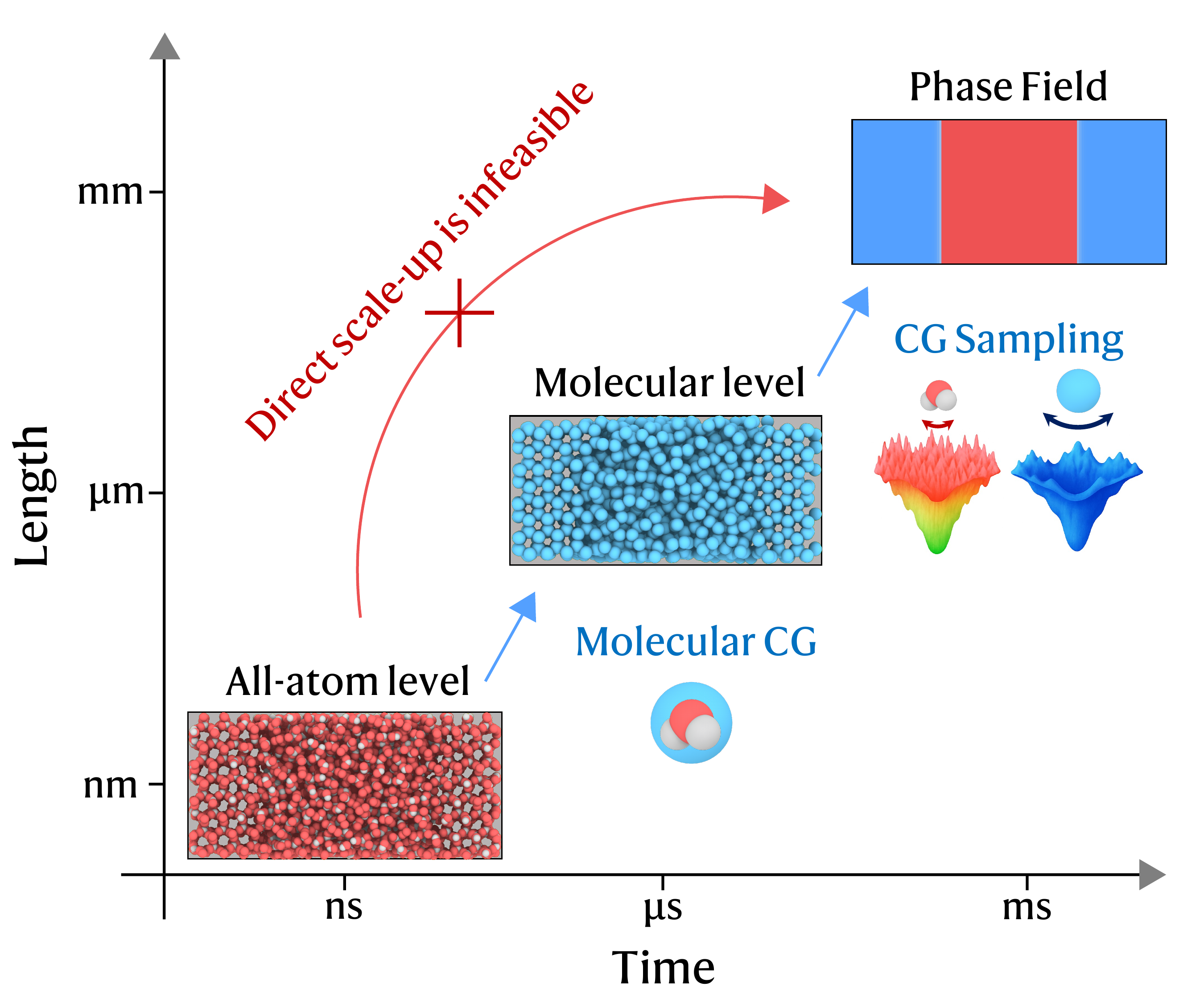}
\caption{\label{fig:fig1} Schematic illustration of the hierarchical coarse-graining (CG) strategy for constructing a mesoscopic representation of the water-ice interface directly from atomistic simulations. To derive microscopic phase-field dynamics from fully atomistic data, the approach involves two key steps: (1) molecular coarse-graining and (2) CG-level sampling.}
\end{figure}

To overcome the limitations of conventional phenomenological phase-field models, we propose a bottom-up framework for constructing a microscopic phase-field model \cite{jin2025field}. Deriving a phase-field model that spans a wide range of length and time scales directly from fully atomistic simulations (typically limited to nanometers and nanoseconds) is both technically challenging and often impractical, as it can lead to overfitting due to the sheer number of degrees of freedom involved. Rather than attempting to reduce millions of atomistic degrees of freedom into a small set of mesoscopic parameters in a single step, our framework addresses this challenge by introducing a hierarchical coarse-graining strategy, as illustrated in Fig. \ref{fig:fig1}. In this framework, we first construct a molecular CG model by integrating out less important atomistic details, here down to the center-of-mass level. The construction of the phase-field model from this intermediate molecular CG model is expected to significantly reduce the challenges associated with developing bottom-up phase-field models. At the reduced level, we then apply enhanced sampling techniques to sample and parametrize the phase-field free energy functional. 

To showcase our approach, water nucleation is an ideal system due to both its significance in fields such as atmospheric chemistry \cite{rosenfeld2000suppression} as well as the fact that the process is difficult to describe over long length and time scales using molecular dynamics simulations \cite{KELTON199175,matsumoto2002molecular}. Additionally, water nucleation serves as a fundamental motif for understanding complex solidification processes that are often beyond the reach of detailed microscopic understanding \cite{sosso2016crystal}. To demonstrate the feasibility of the proposed approach, our initial focus is on capturing crystal growth at the water-ice interface; specifically, we consider undercooled conditions to facilitate crystal growth \cite{Angell1982}. Phase coexistence allows us to determine the distinct free energies of each phase concurrently [Eq. (\ref{eq:1})].

\begin{figure*}
\includegraphics[width=2\columnwidth]{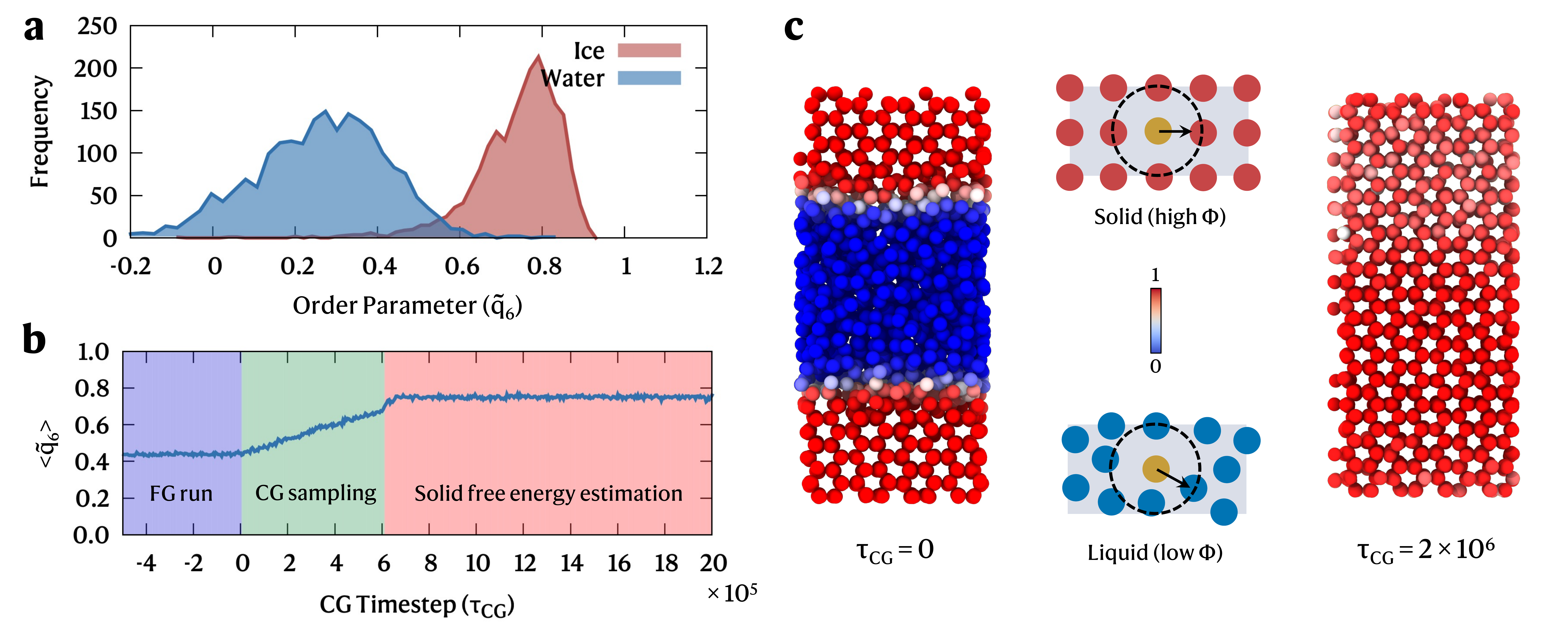}
\caption{\label{fig:fig2} Systematic construction of a molecular OP for ice growth at the ice-water interface ($31.50\,\mathrm{\AA}\,\times\,29.84\,\mathrm{\AA}\times\,68.58\,\mathrm{\AA}$). (a) The local bond OP $\tilde{q}_6$ can distinguish ice (red) from water (blue), demonstrating that $\phi$ can be systematically derived from $\tilde{q}_6$. (b) Time evolution of $\langle \tilde{q}_6\rangle$ during FG and CG simulations. The CG simulation (green) enables enhanced sampling of ice-like configurations, increasing $\tilde{q}_6$ from around 0.4 (FG, blue) to 0.8. After crystallization, an additional CG run (red) was utilized to estimate the absolute free energy density of the solid phase. (c) $\tilde{q}_6$ distribution in the initial configuration ($\tau_\mathrm{CG}=0$) and the final configuration ($\tau_\mathrm{CG}=2\times10^6$) from the molecular CG simulation.}
\end{figure*}

\subsection*{Molecular Order Parameter}
In conventional phase-field modeling of crystal growth, the phase-field $\phi$ is typically introduced as a phenomenological \textit{non-conserved} field variable, which, in the case of water-ice nucleation, takes on 0 for water and 1 for ice. Due to the non-conserved nature of $\phi$, its phase dynamics should follow the Allen--Cahn dynamics \cite{allen1979microscopic}, which corresponds to Model A dynamics in the classification of Hohenberg and Halperin \cite{hohenberg1977theory}
\begin{equation} \label{eq:modela}
    \frac{\partial \phi}{\partial t} = -M_\phi \frac{\delta \mathcal{F}}{\delta \phi},
\end{equation}
where $\mathcal{F}$ is the phenomenological free energy functional defined in Eq. (\ref{eq:1}), and $M_\phi$ denotes the mobility of $\phi$. As $\phi$ is defined only at the continuum field level, it lacks a direct microscopic interpretation, and its evolution and physical meaning are imposed phenomenologically rather than derived from molecular-level physics. Nevertheless, to derive a bottom-up phase-field model with a microscopically informed free energy functional [Eq. (\ref{eq:1})] for eventual use with Eq. (\ref{eq:modela}), one must first construct a ``microscopic'' analog $\{\phi_I\}_I$, defined at the level of individual CG molecules $I$.

For modeling crystal growth, we draw inspiration from microscopic order parameters (OPs) widely used in molecular simulation of crystallization \cite{russo2012microscopic,sosso2016crystal}. In particular, bond OPs are capable of quantifying local orientational symmetry, which undergoes significant changes during crystallization \cite{steinhardt1983bond}. Ice Ih, for example, exhibits a highly ordered, tetrahedrally coordinated structure, leading to bond OPs higher than those of liquid water, which has reduced orientational order [Fig. \ref{fig:fig2}(c)]. Hence, we construct our molecular OP for CG particle $I$ from the local $q_6(I)$ OP (see \textit{Models and Methods} for its definition). While $q_6$ is a robust indicator of local crystalline order, it is susceptible to thermal fluctuations and has been shown to be inadequate in distinguishing between water and ice phases \cite{lechner2008accurate}. To overcome this limitation, we adopt the method proposed by ten Wolde, Ruiz-Montero, and Frenkel, where the local OP is defined as the dot product of the normalized $q_6$ vectors of neighboring particles $I$ and $J$, $\tilde{q}_6(I)$ (see \textit{Models and Methods}), providing a more robust measure of ``crystallinity'' \cite{rein1996numerical}.

To assess the accuracy of this OP for modeling crystal growth, we apply it to trajectories of both bulk water and ice. Figure \ref{fig:fig2}(a) shows that the liquid distribution of $\tilde{q}_6$ ranges from 0 to 0.5, while the ice distribution is primarily localized between 0.5 and 1.0. This bimodal distribution confirms that $\tilde{q}_6$ effectively differentiates between ice and water along the CG trajectories. This approach serves as the basis for our microscopic phase-field variable $\phi_I$.
 
\section*{Results and Discussions}
\subsection*{CG Sampling}
With a well-defined set of molecular-level phase-field variables $\{ \phi_I\}_I$, we proceed to derive the microscopic free energy functional $\mathcal{F}_I (\phi_I)$ that constitutes the total free energy $\mathcal{F}=\sum_I \mathcal{F}_I (\phi_I)$. Each local contribution is assumed to follow a Ginzburg--Landau-type expression, yielding the following form for the total free energy functional:
\begin{align} \label{eq:ucg}
    \mathcal{F} = \sum_I \Biggl( p(\phi_I)f_s+(1-p(\phi_I))f_l + f_w q(\phi_I)+\frac{f_\epsilon^2}{2} |\nabla_I \phi_I|^2 \Biggr),
\end{align}
where $p(\phi_I)$ and $q(\phi_I)$ denote the local probabilities that CG particle $I$ resides in the solid phase or at an interface, respectively. While this Ginzburg--Landau form is adopted here, future work will focus on refining this formulation by deriving the analytical expression through bottom-up coarse-graining formalisms, such as the internal state formalism \cite{dama2017theory,jin2018ultra}.

From Eq. (\ref{eq:ucg}), we first focus on the per-particle bulk free energy functional $\mathcal{F}_I^\mathrm{bulk}(\phi_I):=p(\phi_I)f_s+(1-p(\phi_I))f_l + f_w q(\phi_I)$, which does not involve the gradient term. We immediately notice that $\mathcal{F}_I^\mathrm{bulk}$ readily corresponds to the free energy profile along $\phi_I$, which can be alternatively estimated from enhanced sampling calculations. 

In studies of crystallization and nucleation, rare event sampling techniques, such as umbrella sampling \cite{torrie1977nonphysical} and metadynamics \cite{laio2002escaping,barducci2011metadynamics}, are widely used to overcome the large free energy barriers associated with nucleation in water and other systems \cite{giberti2015metadynamics,sosso2016crystal}. However, in most rare event sampling simulations, a global OP representing the entire system is typically used as the reaction coordinate \cite{van1992computer,ten1995numerical,ten1996simulation,rein1996numerical,trudu2006freezing, quigley2009metadynamics,palmer2013liquid}. In contrast, local OPs are not directly employed to drive the sampling process \cite{eslami2017local}; instead, they are generally computed afterward to identify and characterize the critical nucleus \cite{van1992computer,ten1995numerical,ten1996simulation,auer2001prediction}. Once the nucleus is identified, an effective OP based on heuristically defined clusters is often introduced to bias the simulation \cite{tribello2017analyzing}, and the corresponding free energy is interpreted as the barrier for the formation of a critical crystal nucleus \cite{auer2001prediction}. In our study, since nucleation occurs at an interface rather than under homogeneous bulk conditions, we bypass the need to bias local phase-field parameters by directly performing CG simulations. By removing irrelevant degrees of freedom while retaining essential three-body interactions at the center-of-mass level \cite{larini2010multiscale} (see \textit{Models and Methods}), the bottom-up CG simulation of the water-ice interface is able to capture the spontaneous crystallization into the ice phase \cite{jin2021new1, jin2021new2}. As a result, unbiased CG simulations offer a direct and effective means of sampling the OP space. Figure \ref{fig:fig2}(b) supports this observation, showing that $\tilde{q}_6$ can be treated as a microscopic, non-conserved OP for ice growth, increasing from 0.5 (indicative of ice-water mixture) to nearly 1 over $6\times10^5$ CG MD timesteps ($\tau_\mathrm{CG}$).

Building upon the observation that the unbiased CG simulation effectively samples the $\tilde{q}_6$ space during ice growth, we leveraged the CG trajectory to derive the corresponding PMF, which serves as the bulk free energy functional $\mathcal{F}_I^\mathrm{bulk}$ for our system. From the transition region of the CG trajectory [see Fig. \ref{fig:fig2}(b)], we performed a histogram-based analysis to statistically combine data from different points along the OP \cite{ferrenberg1989optimized}. The Multistate Bennett Acceptance Ratio (MBAR) method was employed to properly weight configurations from across the transition \cite{shirts2008statistically}, resulting in an accurate estimation of $\mathcal{F}_I^\mathrm{bulk}$ from the unbiased CG simulation. 
\begin{figure}
\includegraphics[width=1.0\columnwidth]{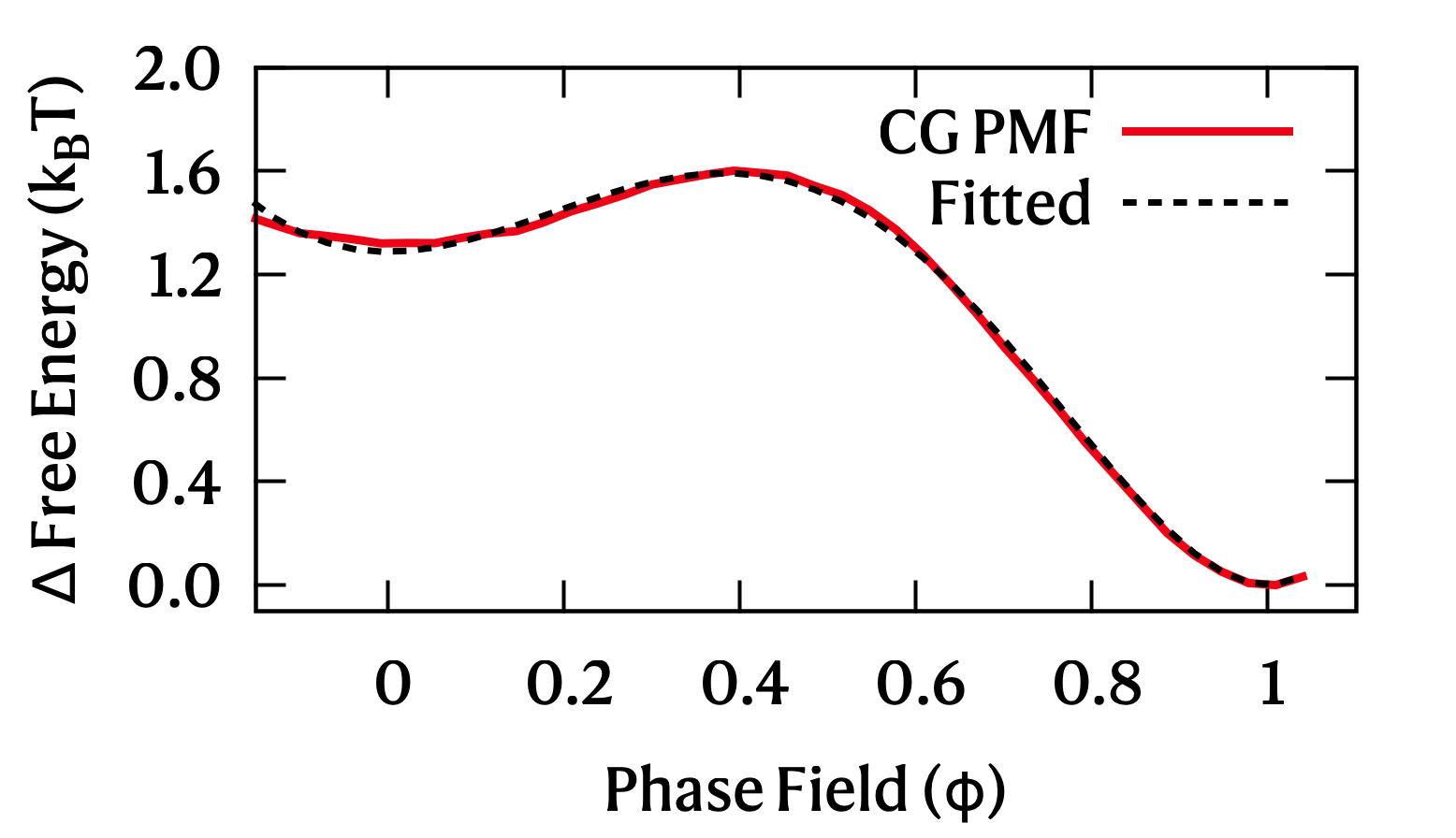}
\caption{\label{fig:fig3} Microscopic bulk free energy landscape underlying ice growth at the interface, expressed in terms of the OP $\phi$. The CG PMF (solid red line) is obtained from histogram analysis using MBAR and represents a relative free energy, shifted such that the solid-phase value is zero. The dashed line shows a fit to the CG PMF using the relative free energy $(1-p(\phi))(f_l-f_s) + q(\phi)f_w$, from which the interpolating polynomials $p(\phi)$ and $q(\phi)$ are determined, cf. Fig. \ref{fig:fig4}.}
\end{figure}

From the histogram analysis, the minimum and maximum values of $\tilde{q}_6$ were determined as $\mathrm{min}(\tilde{q}_6)=0.301$ and $\mathrm{max}(\tilde{q}_6)=0.814$, respectively, consistent with the distribution shown in Fig. \ref{fig:fig2}(a). To maintain consistency with the definition of a phase-field $\phi$, which is stable at $\phi=0$ and $\phi=1$ corresponding to local minima, we finally construct the microscopic phase-field $\phi$ by rescaling $\tilde{q}_6$:
\begin{equation} \label{eq:phi}
\phi = \frac{\tilde{q}_6-\mathrm{min}(\tilde{q}_6)}{\mathrm{max}(\tilde{q}_6)-\mathrm{min}(\tilde{q}_6)}.
\end{equation}

Combined with a histogram analysis, $\phi$ in Eq. (\ref{eq:phi}) is entirely derived from the microscopic distribution of $\tilde{q}_6$ when naturally imposing the condition $\partial \mathcal{F}_I^\mathrm{bulk}/\partial \phi=0$ at $\phi=0$ and $\phi=1$, which are required for the interpolating polynomials. Figure \ref{fig:fig3} depicts the resultant PMF as a function of the microscopic phase-field $\phi$ from the CG sampling. The free energy pathway, with a stable ice phase ($\phi=1$) and a positive energy barrier, is indicative of nucleation and exhibits the characteristic shape expected from phase-field models of undercooled melt \cite{granasy2019phase}. Notably, our free energy profile shows good agreement with the independently obtained landscape from the well-established mW model \cite{molinero2009water}, underscoring the consistency of our approach. Given this well-defined free energy landscape derived from the molecular level, the phase-field framework now enables us to simulate the growth process at much larger length and time scales, extending into the micrometer regime that remains inaccessible to direct molecular simulations.

\subsection*{Limitation of Conventional Phase-Field Model} 
\begin{figure}
\includegraphics[width=1.0\columnwidth]{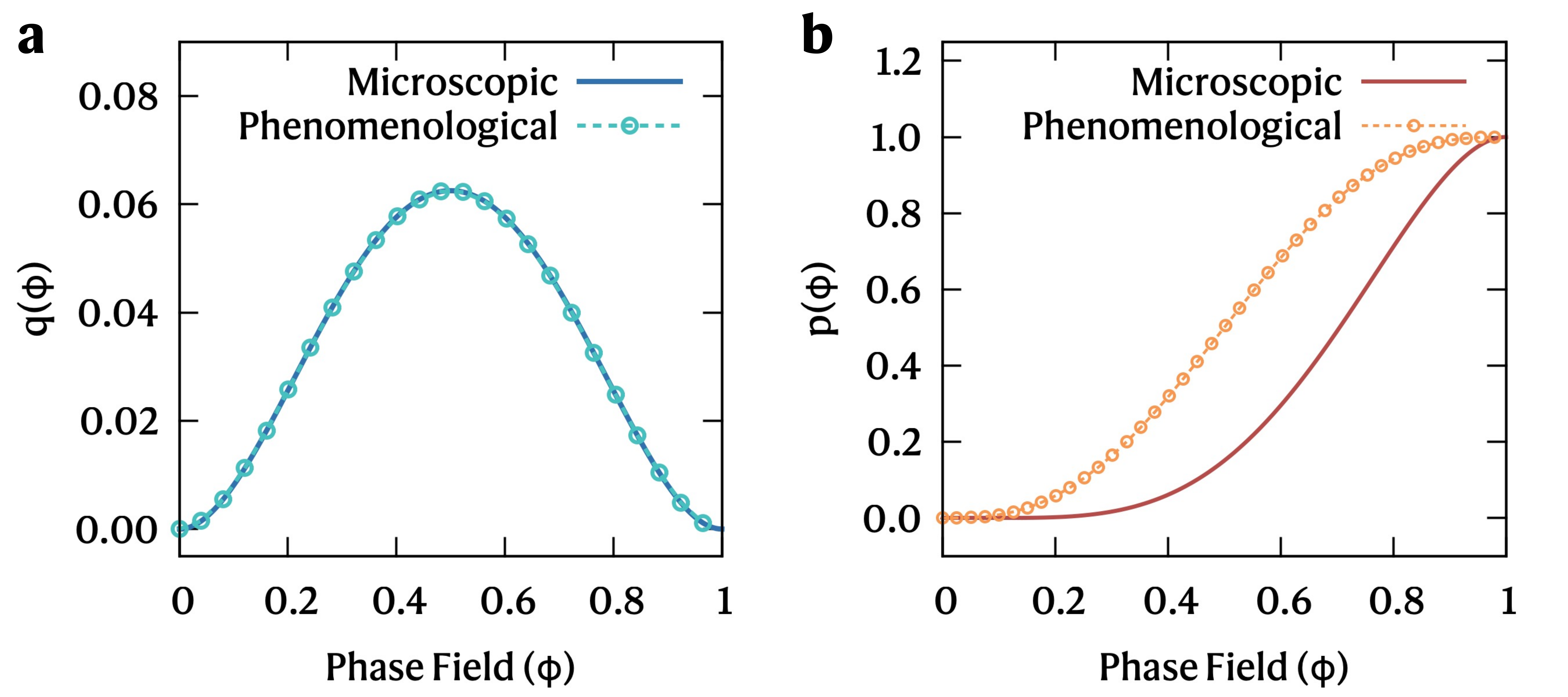}
\caption{\label{fig:fig4} Microscopically derived interpolating polynomials (solid lines): (a) $q(\phi)$ for the double-well free energy and (b) $p(\phi)$ for the chemical free energy, which exhibit slight deviations from the conventionally used phenomenological forms (dots). These microscopic polynomials were obtained by fitting the bulk free energy $F_I^\mathrm{bulk}$ to the CG PMF derived from a histogram analysis of the CG sampling (Fig. \ref{fig:fig3}).}
\end{figure}

We note that the microscopic PMF obtained as a function of the parametrized $\phi$ within our framework retains full microscopic information without invoking any approximation or assuming the Ginzburg--Landau form [Eq. (\ref{eq:ucg})]. To evaluate the validity of the conventional phase-field approximation, we derive the free energy densities ($f_s,\,f_l,\,f_w,$ and $f_\epsilon$) and the interpolating polynomials $p(\phi),\,q(\phi)$ directly from the computed PMF. Since the PMF obtained by histogram analysis represents a relative free energy with an arbitrary constant shift (zeroed at $\phi=1$), the absolute values of the free energy densities must be determined independently. Before extracting the free energy densities, we first determine the interpolating polynomials by fitting $(1-p(\phi))(f_l-f_s)+q(\phi)f_w$ to the PMF profile in Fig. \ref{fig:fig3}. To ensure thermodynamic consistency, the interpolating polynomials must satisfy the boundary conditions: $p(0)=p'(0)=p'(1)=0,$ $ p(1)=1,$ and $q(0)=q(1)=q'(0)=q'(1)=0$. We identified the minimal-degree polynomials satisfying these constraints through nonlinear least‐squares minimization. The resulting polynomials for microscopic phase-field models are $p(\phi)=\phi^2(0.3407-3.4221\phi+10.8220\phi^2-6.7406\phi^3)$ and $q(\phi)=\phi^2(\phi - 1)^2$ with $f_l-f_s=1.288 k_B T$ and $f_w=6.605 k_BT$, as shown in Fig. \ref{fig:fig4}.

Notably, we find that the optimized $q(\phi)$ is identical to the \textit{ad hoc} double-well polynomial $\bar{q}(\phi)$ with symmetry at $\phi=1/2$. As the symmetric double-well polynomial is required to maintain a constant driving force across the diffuse interface and prevent non-physical deformation of the interface profile \cite{steinbach2009phase, moelans2011quantitative}, our findings suggest that a microscopic phase-field variable designed from the local bond OP can provide a physical basis for the functional forms often assumed in continuum-level phase-field models of solidification. While including higher order terms for $q(\phi)$ could improve the fitting of the PMF, this agreement suggests that the \textit{ad hoc} $\bar{q}(\phi)$ polynomial can reasonably approximate the double-well barrier for ice growth. However, we observe notable differences between $p(\phi)$ and $\bar{p}(\phi)$. While $p(\phi)$ interpolates between water and ice by satisfying $p(0)=0$ and $p(1)=1$, the general profile of $p(\phi)$ between $\phi=0$ and $1$ is noticeably different from the \textit{ad hoc} profile. We attribute this deviation to molecular-level details. Consequently, Fig. \ref{fig:fig4} confirms that the interpolating polynomials should generally differ from the \textit{ad hoc} polynomials due to the microscopic nature of different systems, where our approach can improve the limitation of the classical phase-field free energy functional by incorporating system-dependent corrections into $\bar{p}(\phi)$ and $\bar{q}(\phi)$. Later, we will demonstrate that these microscopically informed polynomials impart an accurate description of the growth process. 

\subsection*{Full Parametrization of Microscopic Phase-Field Model}
Having parametrized the microscopic interpolating polynomials, we now estimate the absolute free energy densities in the free energy functional from CG sampling. While directly determining free energy quantities from atomistic simulations, in principle, requires computationally expensive free energy sampling, 
the first coarse-graining step in our approach inherently avoids this limitation. 

Because the effective CG interactions are free energy-based quantities, i.e., many-body CG PMFs \cite{jin2022bottom}, we can directly estimate free energy densities by evaluating the renormalized potential energy at the CG level. Once the system crystallizes into ice after approximately $\tau_{\mathrm{CG}}=6\times10^5$ timesteps [red area in Fig. \ref{fig:fig2}(b)], the gradient contribution, $f_\epsilon^2/2|\nabla_I\phi_I|^2$, becomes negligible due to the relatively small spatial fluctuations in $\phi_I$. Under this condition, the total potential energy $V_\mathrm{tot}$ is assumed to be dominated by the bulk solid contribution, allowing us to estimate the solid-phase free energy $f_s$ by computing $V_\mathrm{tot} / \sum_I p(\phi_I)$. Averaging over $3 \times 10^5$ snapshots collected after crystallization, we obtain $f_s = -8.677$ kcal/mol and $f_l = -8.165$ kcal/mol.

Having determined the bulk free energy densities, we next estimate the gradient free energy density during the crystallization sampled by the CG simulation [green area in Fig. \ref{fig:fig2}(b)]. Since $\tilde{q}_6$ involves nonlocal spherical-harmonic correlations, deriving its spatial derivatives in closed form is extremely complex and prone to numerical instability. Hence, we compute $|\nabla_I \phi_I|^2$ in $\mathcal{F}_I$ by finite differences: each particle $I$ is displaced by $\pm h$ along $x,\,y,$ and $z$ to form central differences ($h=10^{-3}$ $\mathrm{\AA}$), which are then converted analytically to $\nabla_I \phi_I$. Finally, we subtract the bulk free energy contribution (using $f_s,\,f_l,$ and $f_w$) from the CG potential energy sampled during the transition and divide the remainder by $\sum_I |\nabla_I \phi_I|^2$ to obtain $f_\epsilon=1.416$ kcal/mol (\textit{SI Appendix}, Fig. S4). We emphasize that our CG approach correctly captures the underlying physics of solidification by satisfying ${f}_s-{f}_l<0,\,{f}_w>0,$ and ${f}_\epsilon>0$ from the parametrization.

\subsection*{Bottom-Up Phase-Field Model}

Finally, we extend the microscopic CG free energy functional to mesoscopic phase-field thermodynamics to gauge the model's ability to reproduce nucleation. The CG bulk free energy densities (${f}_s,\,{f}_l,$ and ${f}_w$ in kcal/mol) are directly scaled to the mesoscopic energy densities ($\mathbb{f}$ in J/m$^3$) by averaging the CG energetics over the molecular volume. The thermodynamic consistency of our method enables predictions of whether the system exhibits one phase ($\mathbb{f}_l<\mathbb{f}_s$) or two phases ($\mathbb{f}_l>\mathbb{f}_s$) at different temperatures. At the mesoscale, we obtained a positive energy barrier, $\mathbb{f}_w = 2.832 \times 10^5 \,\mathrm{J/m^3}$, and a stable solid phase, $\mathbb{f}_l-\mathbb{f}_s=5.522 \times10^4 \,\mathrm{J/m^3}$, indicative of nucleation.

\begin{figure*}
\includegraphics[width=2\columnwidth]{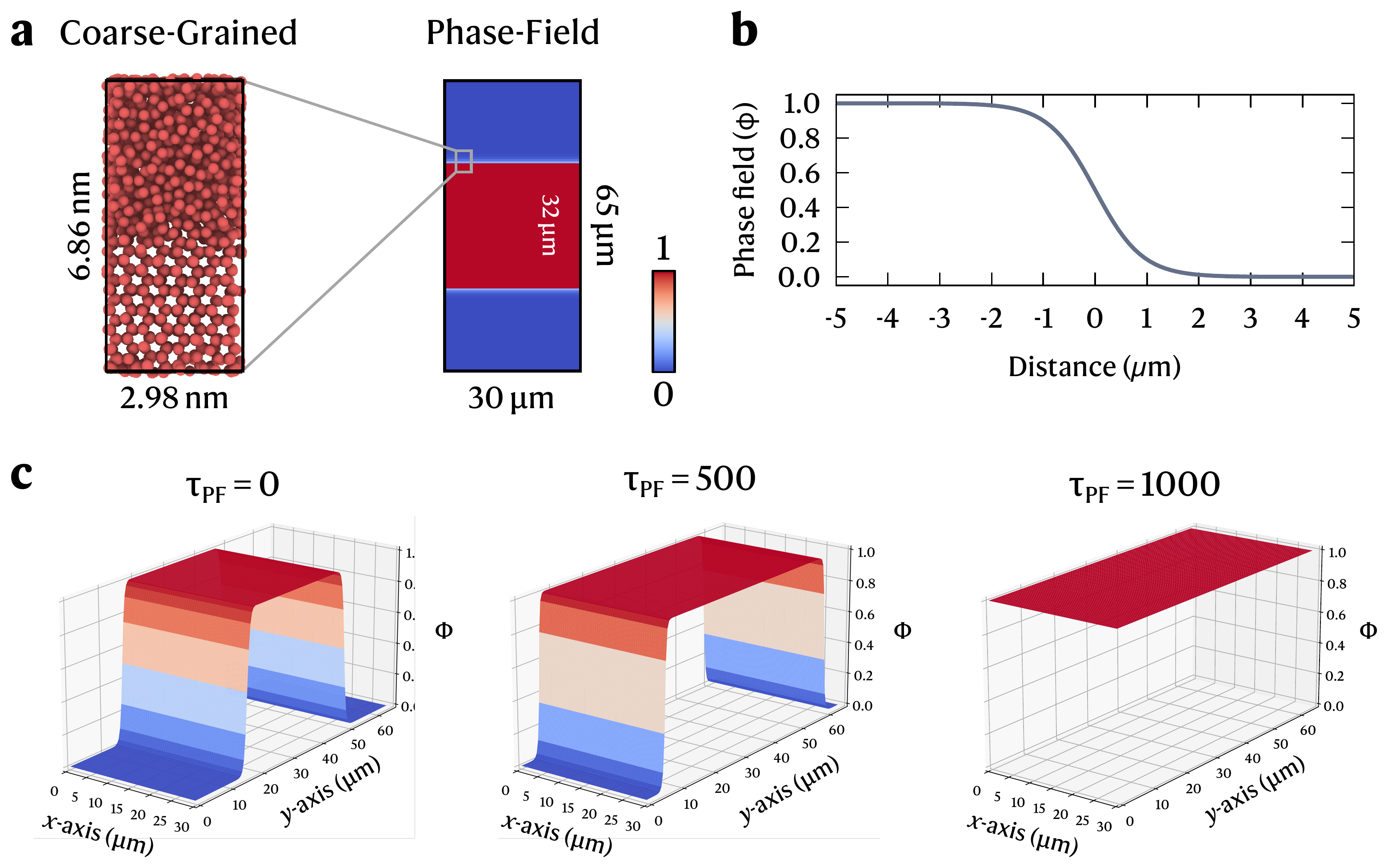}
\caption{\label{fig:phasefield} Microscopically-derived phase-field model for ice growth. (a) Scaling up the molecular CG interface to the phase-field level. (b) Equilibrium profile of phase-field $\phi$ from the center of interface. (c) Time evolution of $\phi$ driven by microscopically-derived free energy functional using \textit{Model A dynamics} [Eq. (2)], where $\tau_{PF}$ is the phase-field timestep ($\phi=1$: ice, $0$: water).}
\end{figure*}

Unlike bulk quantities, directly mapping the interface energy density $\bar{f}_\epsilon$ to the mesoscopic interface term, $\mathbb{f}_\epsilon:=\mathbb{e}^2/2$, is not straightforward due to the significant scale difference between molecular-level length scales ($<\mathrm{nm}$) and the mesoscopic interface width $\delta$ ($\sim \mathrm{\mu m}$). Nevertheless, assuming steady-state interface growth, the gradient energy coefficient $\mathbb{e}$ can be determined from mesoscopic physical properties \cite{warren1995prediction,takaki2005phase}. In doing so, we first build a phase-field level mesoscopic interface (30 $\mathrm{\mu m} \times$65 $\mathrm{\mu m}$) from the atomistic level (2.98 nm$\times$6.86 nm) with a similar aspect ratio [Fig. \ref{fig:phasefield}(a)]. We note that it is computationally prohibitive for particle-based CG models to reach beyond micrometer regimes, but we can achieve nearly a $10^8$-fold scale-up with microscopic phase-field models. 

To discretize the phase-field system onto numerical grids, we chose a lattice size of $\Delta x = 0.5\, \mathrm{\mu m}$, following typical values for pure materials \cite{wheeler1993computation} to achieve an interface thickness of $\delta = 4\Delta x$, covering $\lambda=0.1<\phi<1-\lambda$. Under the steady-state approximation \cite{takaki2005phase}, $\mathbb{e}$ is related to $\delta$ via the interface region parameter $b=2\mathrm{tanh}^{-1}(1-2\lambda)$ such that $\mathbb{e} = \sqrt{3\delta\gamma/b}$, where the interface energy $\gamma$ between phases is linked to $\mathbb{f}_w$ with $\gamma = (\mathbb{f}_w \delta)/(6b)$ (\textit{SI Appendix}, Eqs. (3)--(7)). Together, the mesoscopic interface term is estimated as $\mathbb{e} = 3.4249 \times10^{-4}\, \mathrm{J^{1/2}/m^{1/2}}$. Having determined $\mathbb{f}_w$ and $\mathbb{f}_\epsilon$, the initial condition $\phi_{0}(r\Delta x)$ from the center of the interface follows the equilibrium profile by solving $\partial\phi/\partial t = 0$ [Eq. \eqref{eq:phi0}], resulting in an interface thickness of approximately $4\,\mathrm{\mu m}$, as shown in Fig. \ref{fig:phasefield}(b).

Our approach establishes a well-defined and useful segue from the description of the molecular system to the field-level simulation of Model A dynamics. Equation (\ref{eq:modela}) was propagated using the microscopically derived free energy functional $\mathcal{F}_{PF} = \int_V dV \left( p(\phi)\mathbb{f}_s+(1-p(\phi))\mathbb{f}_l+\mathbb{f}_w q(\phi)+\mathbb{f}_\epsilon |\nabla\phi|^2\right)$ and the phase-field mobility $M_\phi = 0.366 \mathrm{m^3\cdot J^{-1}s^{-1}}$ based on conventional settings with an interfacial mobility of $10^{-6} \mathrm{m^4\cdot J^{-1}s^{-1}}$ (see \textit{Discussion} in \textit{SI Appendix}), which determines the effective timestep $\tau_{PF}$ \cite{mobility}. Remarkably, Fig. \ref{fig:phasefield}(c) shows that our \textit{ab initio} phase-field model successfully recapitulates the fully crystallized ice phase ($\tau_{PF}=1000$), evolving from the ice-water interface ($\tau_{PF}=0$), consistent with molecular-level phenomena. This finding underscores that the proposed approach can accurately predict crystal growth at the mesoscale based entirely on detailed microscopic interactions, without relying on experimental data. 

\section*{Conclusion}

We present a novel multiscale framework for deriving microscopic phase-field models grounded entirely in microscopic dynamics by leveraging bottom-up coarse-graining and enhanced sampling techniques. This microscopic-mesoscopic link, spanning eight orders of magnitude in spatiotemporal scales, is achieved through a hierarchical coarse-graining approach. Our framework requires only microscopic simulations, eliminating the need for \textit{ad hoc} thermodynamic parameters. By incorporating system-specific statistics, our findings offer new insights into improving conventional phase-field models, particularly the interpolating polynomials, to more accurately represent complex phase dynamics.

In constructing the microscopic phase-field model, enhanced sampling allows derivation of the microscopic free energy functional. As the first step in this approach, our work uses CG MD simulations to accelerate sampling in the OP space. Nevertheless, existing enhanced sampling techniques, such as umbrella sampling or metadynamics \cite{laio2002escaping}, can be seamlessly integrated into our framework to facilitate efficient sampling of the free energy functionals underlying complex phase dynamics beyond solidification behavior in atomistic or CG systems. Because the framework developed in this work is general by design, such combined approaches will further enhance the applicability and efficiency of microscopic phase-field models. This integration will be explored in future work. Especially, since our methodology is readily generalizable to any arbitrary phase dynamics, such as multi-phase-field \cite{steinbach1996phase} or phase-field crystal models \cite{elder2007phase}, we anticipate that its impact will extend beyond the scope of rigorous coarse-graining and enable the exploration of new classes of mesoscopic materials and phenomena.

\matmethods{
\subsection*{Molecular Order Parameter}
Following Steinhardt \textit{et al.}, the local $q_6(I)$ OP for the CG water particle \textit{I} is defined as \cite{steinhardt1983bond}
\begin{equation} \label{eq:q6}
    q_6(I) = \left(\frac{4\pi}{13} \sum_{m=-6}^6 |q_{6m}(I)|^2\right)^{1/2},
\end{equation}
where $q_{6m}(I)$ is defined as $\sum_{J\in N_b(I)} Y_6^m (\theta_{IJ},\phi_{IJ}
)/{N_b(I)}$ with $Y_{6}^m$ being the $6m$-th spherical harmonic, and $\theta_{IJ}$ and $\phi_{IJ}$ as the angles associated with vector $\vec{R}_{IJ}$. The number of neighbors of particle $I$ is $N_b(I)$, defined by the switching function $N_b(I)=\sum_J \sigma(R_{IJ}),$ where the cutoff function is given as $\sigma(R) = [1-(R/3.5\,\mathrm{\AA})^6]/[1-(R/3.5\,\mathrm{\AA})^{12}]$. Next, we construct the vector $\mathbf{q}_6(I)=(q_{6,-6}(I),q_{6,-5}(I),\cdots,q_{6,6}(I))$ and finally compute $\tilde{q}_6(I)$ for molecule \textit{I} as 
\begin{equation}
\tilde{q}_6(I) = \frac{1}{N_b(I)} \sum_{J\in N_b(I)} \frac{\mathbf{q}_6(I)\cdot \mathbf{q}_6^\ast(J)}{|\mathbf{q}_6(I)| |\mathbf{q}_6(J)|}.
\end{equation}
By ensuring that CG site $I$ and its local environment share a similar and coherently oriented structure, $\tilde{q}_6(I)$ serves as a better indicator of crystalline regions than $q_6(I)$ \cite{lechner2008accurate}. The global version of Eq. (\ref{eq:q6}), denoted as $Q_6$, is obtained by summing $q_{6m}(I)$ for all CG particles.

\subsection*{Atomistic Simulation Setup}
Following previous CG studies of the ice-water interface \cite{jin2021new1,jin2021new2}, we constructed an interface composed of 1024 water and 1024 ice molecules. The initial water structure was centered within the simulation box, and the size of the simulation box was determined by the experimental density. The liquid box was capped at 512 ice molecules on both top and bottom, and the initial structure was randomized while the ice configuration was initialized with zero dipole moments following Ref. \cite{buch1998simulations} (\textit{SI Appendix}, Fig. S1). After equilibration of each component, the slab configuration was constructed by merging them into a box with dimensions $31.50 \times 29.84 \times 68.58$ (\AA$^3$). The atomistic simulation followed the simulation protocol established in Ref. \cite{jin2021new2} using the TIP4P/Ice force field \cite{vega2005relation} under the undercooled condition $T=249$ K and $P=1$ atm. Both atomistic and CG MD simulations were performed using the large-scale atomic/molecular massively-parallel simulator (LAMMPS) \cite{plimpton1995fast}.

\subsection*{Coarse-Graining Parametrization}
From the atomistic trajectory with configurations $\mathbf{r}^n$, a molecular CG model at the CG configuration $\mathbf{R}^N$ was constructed using a bottom-up approach. Since three-body correlations are essential for describing hydrogen-bonding network and ice structure in water systems \cite{gora2011interaction}, we incorporated effective three-body interactions into the CG Hamiltonian \cite{hankins1970water}: $U(\mathbf{R}^N)=\sum_I\sum_{J\neq I} U^{(2)}(R_{IJ})+\sum_{I}\sum_{J\neq I}\sum_{K>J} U^{(3)} (\theta_{JIK},R_{IJ},R_{IK}),$ 
where $U^{(2)}$ and $U^{(3)}$ denote the two-body and three-body potentials, respectively. While detailed parametrization procedures are described in Refs. \cite{jin2021new1,jin2021new2}, we outline the key elements here with the additional results in \textit{SI Appendix}. Inspired by the monatomic water (mW) model \cite{molinero2009water}, we employed the Stillinger-Weber potential \cite{stillinger1985computer} for the three-body interaction, $U^{(3)}= \lambda_{JIK} \left( \cos \theta_{jik}-\cos \theta_0\right)^2     \allowbreak
\exp\!\left({\gamma_{IJ}}/(R_{IJ}-a_{IJ}) \right)     \allowbreak
\exp\!\left( {\gamma_{IK}}/(R_{IK}-a_{IK})\right),$
where $\gamma_{JIK}$ sets strength of angular interactions, and $a_{IJ}$ and $a_{IK}$ are the cutoff distances. Following the original protocol for CG water \cite {larini2010multiscale}, we set $\gamma_{IJ}=1.2$, $\sigma_{IJ}=1$ \AA, $\epsilon_{JIK}=1.0$ kcal/mol, and $a_{IJ}=3.7$ \AA. 
 The two-body potential $U^{(2)}$ and the angular strength parameter $\lambda_{JIK}$ in $U^{(3)}$ were then variationally optimized using the force-matching approach \cite{larini2010multiscale}, as implemented in the OpenMSCG open-source package \cite{peng2023openmscg}, yielding $\lambda_{JIK}=28.0078$ at 249 K. 
 
\subsection*{Phase-Field Simulation}
We constructed a two-dimensional mesoscopic-level system on a $130\times60$ grid with a spacing of $\Delta x =0.5 \mathrm{\mu m}$, resulting in an interface size of $30\,\mathrm{\mu m} \times 65 \, \mathrm{\mu m}$. The initial condition $\phi_0$ follows the equilibrium profile:
\begin{equation} \label{eq:phi0}
    \phi_{0}(r\Delta x) = \frac{1}{2} \left[ 1- \mathrm{tanh} \left(\sqrt{\frac{\mathbb{f}_w}{\mathbb{f}_\epsilon}}r\Delta x \right)\right]=\frac{1- \mathrm{tanh} \left(0.549r\right)}{2},
\end{equation}
which is superposed at the center of the interface (\textit{SI Appendix}, Fig. S5). We numerically discretized the Allen--Cahn equation using a finite difference method (see \textit{SI Appendix} for computational details). 

\subsection*{Data Availability}
Jupyter Notebook for microscopic phase-field simulations and a parametrization script in the C++ programming language are available on GitHub \cite{github-repo}.

}

\showmatmethods{} 
\acknow{J.J. thanks the Arnold O. Beckman Postdoctoral Fellowship for funding and academic support (http://dx.doi.org/10.13039/100000997).}

\showacknow{} 

\bibliography{manuscript}

\begin{thebibliography}{10}
\bibitem{mobility}
The reader may note that we do not address the parametrization of $m_\phi$ from first principles in this work. however, the relative time can be understood by rescaling the timescale $\tau_{PF}$. a rigorous determination of $m_\phi$ is currently being pursued by the authors using a microscopic time-dependent projection operator approach.
\end{thebibliography}

\end{document}